\documentclass[11pt]{article}
\usepackage{graphicx}
\usepackage{latexsym}
\usepackage{appendix}
\def\CA{{\cal A}}
\def\CB{{\cal B}}

\def\CM{{\cal M}}

\def\CL{{\cal L}}
\def\CN{{\cal N}}
\def\CO{{\cal O}}

\def\centeron#1#2{{\setbox0=\hbox{#1}\setbox1=\hbox{#2}\ifdim
   \wd1>\wd0\kern.48\wd1\kern-.48\wd0\fi
   \copy0\kern-.48\wd0\kern-.48\wd1\copy1\ifdim\wd0>\wd1
   \kern.48\wd0\kern-.48\wd1\fi}}

\newcommand{\beq}{\begin{equation}}
\newcommand{\eeq}{\end{equation}}
\newcommand{\bea}{\begin{eqnarray}}
\newcommand{\eea}{\end{eqnarray}}
\newcommand{\ba}{\begin{array}}
\newcommand{\ea}{\end{array}}

\newcommand{\p}{\partial}
\newcommand{\nn}{\nonumber}

\newcommand{\half}{\frac{1}{2}}

\begin{document}

\hskip3cm

\begin{center}
 \LARGE \bf  AdS Black Hole Solutions\\
in the Extended New Massive Gravity
\end{center}

\vskip2cm

\centerline{\Large Soonkeon Nam\footnote{nam@khu.ac.kr}\,,
~~Jong-Dae Park\footnote{jdpark@khu.ac.kr}\,, ~~Sang-Heon
Yi\footnote{shyi@yonsei.ac.kr}}

\hskip2cm

\begin{quote}
Department of Physics and Research Institute of Basic Science, Kyung
Hee University, Seoul 130-701, Korea$^{1,2}$

Department of Physics, College of Sciences, Yonsei University, Seoul
120-749, Korea$^3$
\end{quote}

\hskip2cm

\vskip2cm

\centerline{\bf Abstract} We have obtained  (warped) AdS black hole
solutions in the three dimensional extended new massive gravity.  We
investigate some properties of black holes and obtain central
charges of the two dimensional dual CFT. To obtain the central
charges, we  use  the relation between entropy and temperature
according to the $AdS/CFT$ dictionary.  For $AdS$ black holes, one
can also use the central charge function formalism which leads to
the same results. \thispagestyle{empty}
\renewcommand{\thefootnote}{\arabic{footnote}}
\setcounter{footnote}{0}
\newpage

%%%%%%%%%%%%%%%%%%%%%%%%%%%%%%%%%%%%%%%%%%%%%%%%%%%%%%%%%%%%%%%%%%%%%%%
\section{Introduction}
%%%%%%%%%%%%%%%%%%%%%%%%%%%%%%%%%%%%%%%%%%%%%%%%%%%%%%%%%%%%%%%%%%%%%%%
%
Recently, there have been some interests in three dimensional
gravity theories containing certain higher curvature terms in the
action, which are initiated by the advent of the new massive
gravity(NMG)\cite{Bergshoeff:2009hq}.  Though there are no
propagating degrees of freedom only with the Einstein-Hilbert term,
the situation becomes different with these higher curvature terms in
the Lagrangian.  It has long been  known that the theory with
gravitational Chern-Simons term~\cite{Deser:1981wh}\cite{Deser:1982}
leads to massive gravitons and also allows black hole
solutions~\cite{Hotta:2008}$\sim$\cite{Clement:2007}. The newly
proposed NMG theory can be thought as the non-linear extension of
Fierz-Pauli massive graviton theory, and it preserves parity
symmetry compared to the topologically massive gravity(TMG). Whereas
it has also been shown that there  are black hole solutions of the
usual BTZ type in the NMG with cosmological constants, there are new
type of black holes~\cite{Bergshoeff:2009aq} and those with the
warped $AdS$ asymptotics~\cite{Clement:2009gq}.

There have been various studies on NMG, which include supersymmetric
extension, its black hole solutions, the central charge of the dual
conformal field theory(CFT), new type black holes,
etc\cite{Bergshoeff:2009hq}\cite{Bergshoeff:2009aq}$\sim$\cite{Townsend:2010}.
In the context of the quantum gravity, it is meaningful to consider
higher curvature corrections. The NMG may be regarded as three
dimensional gravity with curvature square correction terms.
Therefore, it is natural to consider even higher curvature
corrections in the viewpoint of the quantum gravity or the string
theory.

More recently, there are proposals to extend NMG to even higher
curvature theories, one of which is the extension of NMG to $R^3$
terms in the Lagrangian~\cite{Sinha:2010ai}. This is consistent with
the {\it holographic c-theorem}. The other is the extension to the
Born-Infeld type Lagrangian~\cite{Gullu:2010pc}. These may be a good
playground to go toward the quantum gravity and to test the
$AdS/CFT$ correspondence.

These two extensions become identical up to $R^3$ order terms when a
suitable matching among parameters is performed. The linearized
theory, which tells us the properties of gravitons, is not yet
investigated on these theories. Instead of this direction, we
investigate the properties of various $AdS$ black holes in these
theories and dual conformal field theories.

In this $AdS/CFT$ era,  whenever there is a $AdS$ solution in some
gravity theory, it is standard to envisage the existence of the dual
CFT on the boundary and consider the implications of  the gravity
for the dual
CFT~\cite{Maldacena:1997}\cite{Witten:9802}\cite{Witten:9803}. For
the three dimensional gravity theory, it is natural to conjecture
the dual theory as a kind of two dimensional CFT. The first thing
one may try to do is to obtain the central charge of the
hypothetical 2d CFT from gravity theory. This is one reason to study
the entropy, temperature, etc. in these theories in the view point
of $AdS/CFT$ correspondence.

Now, there are several ways to obtain central charges of the dual
CFT.  In this paper, we use mainly the ``Cardy formula'' to obtain
central charges of the dual CFT~\cite{Cardy:86}\cite{Carlip:1998}.
This formula may be thought as the relation between temperature and
entropy. Concretely, by writing the black hole entropy in terms of
the black hole temperature and interpreting those quantities in the
CFT side according to the usual prescription of $AdS_3/CFT_2$
dictionary, the Cardy formula is given by
\[
 S_{BH} = \frac{\pi^2 L}{3}(c_LT_L + c_RT_R)\,.
\]
Since the extended NMG theories considered in this paper are parity
symmetric, the left and the right central charge of the dual CFTs
are identical $c=c_L=c_R$. One of the main results in this paper is
to present central charge, $c$,  of the dual CFTs, explicitly. For
the asymptotically $AdS$ case, there is another way to obtain
central charge, which is developed in~\cite{Brown-Henneaux:86}$\sim$
\cite{Kraus:2005vz}, coined as central charge function formalism. We
propose a slight extension of this formalism to apply to new type
black holes and show that it leads to the value consistent with the
above Cardy formula.

%\cite{Skenderis:1998}\cite{Kraus:1999}

This paper is organized as follows. In section 2, we review briefly
two types of the  extension of NMG proposed recently. One is the
extension of NMG with curvature cubic terms($R^3$-NMG) and the other
is the Born-Infeld type extension of NMG(BI-NMG). In section 3,
after the some explanation of our method to obtain solutions,  we
present various black hole solutions: BTZ black holes, warped $AdS$
black holes and new type black holes.  Various physical quantities,
especially the entropy and central charges of dual CFTs  are
presented.  Since this section is relatively long,  we divide it
into several subsections with some introductory remarks. Our results
are summarized and some future directions are discussed  in section
4. Some calculational details are relegated to the Appendix.

%%%%%%%%%%%%%%%%%%%%%%%%%%%%%%%%%%%%%%%%%%%%%%%%%%%%%%%%%%%%%%%%%%%%%%%%%%
\section{The Extended New Massive Gravity}
%%%%%%%%%%%%%%%%%%%%%%%%%%%%%%%%%%%%%%%%%%%%%%%%%%%%%%%%%%%%%%%%%%%%%%%%%%
%
There are new interests in three dimensional gravity theories with
higher curvature terms. Though there are no propagating degrees of
freedom in three dimensions only with the Einstein-Hilbert term, the
situation becomes different with higher curvature terms in the
Lagrangian.  It has long been  known that the theory with
gravitational Chern-Simons term~\cite{Deser:1981wh}\cite{Deser:1982}
leads to massive graviton and also allows black hole solutions.
Recently, another type of massive graviton theory has been explored
(see for a review~\cite{Bergshoeff:2009}), which is named as new
massive gravity(NMG).

One reason of these developments comes from the implication of these
theories in two dimensional CFT through $AdS/CFT$ correspondence.
One may  ask whether it is possible to extend NMG theory to the
theory with even higher curvature terms. There are recent works to
answer this question. In doing so, we need some guideline to go to
higher curvature terms. If these theories have a string/M-theory
embedding, one can obtain the higher curvature corrections
systematically. However, this embedding is not done yet, and
moreover, it is unclear that there is such embedding at all.
Therefore, it seems desirable to have another way to obtain higher
curvature corrections, for instance, through $AdS/CFT$
correspondence. There is a recent attempt~\cite{Sinha:2010ai}  to
extend  the new massive gravity  by considering the central charge
function and c-theorem, which basically utilize $AdS/CFT$ machinery.
There is another attempt~\cite{Gullu:2010pc} containing an infinite
number of higher curvature terms, which becomes identical with the
previous one up to the curvature cubic order by a suitable parameter
matching.

%%%%%%%%%%%%%%%%%%%%%%%%%%%%%%%%%%%%%%%%%%%%%%%%%%%%%%%%%%%%%%%%%%%%%%
\subsection{The $R^3$  extension of  New Massive Gravity}
%%%%%%%%%%%%%%%%%%%%%%%%%%%%%%%%%%%%%%%%%%%%%%%%%%%%%%%%%%%%%%%%%%%%%%
%
The $R^3$ extension of New Massive Gravity
($R^3$-NMG)~\cite{Sinha:2010ai} is
\begin{equation}\label{NMG}
S =\frac{\eta}{2\kappa^2}\int d^3x \sqrt{-g}\bigg[ \sigma R +
\frac{2}{l^2} + \frac{1}{m^2}K + \frac{\xi}{12\mu^4}K'\bigg]\,,
\end{equation}
where $2\kappa^2= 16\pi G$ is three dimensional Newton's constant
and $\eta$,  $\sigma$ and $\xi$ take $1$ or $-1$. Here, $K$ and $K'$
are defined by
\begin{eqnarray}
 K &=& R_{\mu\nu}R^{\mu\nu} -\frac{3}{8}R^2\,, \nn \\
 K' &=&17R^3 -72R_{\mu\nu}R^{\mu\nu}R
       + 64R_{\mu}^{\, \nu}R_{\nu}^{\, \rho}R_{\rho}^{\, \mu}\,. \nn
\end{eqnarray}
At the level of equations of motion, the overall sign $ \eta$ of the
action is meaningless, but it has some important consequence in the
black hole entropy and in the positivity of the central charge of
dual CFT. So, we allow the overall sign choice  $\eta$. Our
convention is such that $\kappa^2, m^2 , \mu^2$ are always positive,
but cosmological constant, $1/l^2$ may be positive or negative.

Note that the above  $K$ and $K'$ satisfy the following interesting
relations
\[
  g_{\mu\nu}\frac{\p K}{\p R_{\mu\nu}}
     = -\frac{1}{4} R\,, \qquad
  g_{\mu\nu}\frac{\p K'}{\p R_{\mu\nu}}
     = -24 K\,.
\]

The equations of motion for the $R^3$-NMG are
\begin{equation}
 \sigma G_{\mu\nu} - \frac{1}{l^2}g_{\mu\nu} + \frac{1}{2m^2}K_{\mu\nu}
        - \frac{\xi}{12\mu^4}K'_{\mu\nu} =0\,,
\end{equation}
where
\begin{eqnarray}
 K_{\mu\nu} &=& g_{\mu\nu}\Big(3R_{\alpha\beta}R^{\alpha\beta}-\frac{13}{8}R^2\Big)
                + \frac{9}{2}RR_{\mu\nu} -8R_{\mu\alpha}R^{\alpha}_{\nu}
                + \half\Big(4\nabla^2R_{\mu\nu}-\nabla_{\mu}\nabla_{\nu}R
                -g_{\mu\nu}\nabla^2R\Big)\,, \nn \\
 K'_{\mu\nu} &=& 17\Big[-3R^2R_{\mu\nu} +3\nabla_{\mu}\nabla_{\nu}R^2
                +\half g_{\mu\nu}R^3-3g_{\mu\nu}\nabla^2R^2\Big] \nn \\
 && -72\Big[-2RR_{\mu\alpha}R^{\alpha}_{\nu} -R_{\alpha\beta}R^{\alpha\beta}R_{\mu\nu}
                -\nabla^2(RR_{\mu\nu}) +\nabla_{\mu}\nabla_{\nu}(R_{\alpha\beta}R^{\alpha\beta})
                +2\nabla_{\alpha}\nabla_{(\mu}(R_{\nu)}^{\alpha}R) \nn \\
 && + \half g_{\mu\nu}R_{\alpha\beta}R^{\alpha\beta}R
                -g_{\mu\nu}\nabla_{\alpha}\nabla_{\beta}(RR^{\alpha\beta})
                -g_{\mu\nu}\nabla^2(R_{\alpha\beta}R^{\alpha\beta})\Big] \nn \\
 && + 64\Big[ -3R_{\mu}^{\, \rho}R_{\rho}^{\, \sigma}R_{\sigma\nu}
                -\frac{3}{2}\nabla^2(R_{\mu\alpha}R^{\alpha}_{\nu})
                +3\nabla_{\alpha}\nabla_{(\mu}(R_{\nu)}^{\beta}R_{\beta}^{\alpha}) \nn \\
 &&  + \half g_{\mu\nu}R_{\alpha\beta}R^{\alpha\beta}
                -\frac{3}{2}g_{\mu\nu}\nabla_{\alpha}\nabla_{\beta}
                (R^{\alpha}_{\rho}R^{\rho\beta})\Big]\,.  \nn
\end{eqnarray}

%%%%%%%%%%%%%%%%%%%%%%%%%%%%%%%%%%%%%%%%%%%%%%%%%%%%%%%%%%%%%%%%%%%%%%%%%%%
\subsection{The Born-Infeld Extension of New Massive Gravity}
%%%%%%%%%%%%%%%%%%%%%%%%%%%%%%%%%%%%%%%%%%%%%%%%%%%%%%%%%%%%%%%%%%%%%%%%%%%
%
The Born-Infeld extension of the new massive gravity
(BI-NMG)~\cite{Gullu:2010pc} is also introduced,  whose action is
\begin{equation}\label{BI-action}
 S = - \eta\frac{2m^2}{\kappa^2}\int d^3x \sqrt{-g}
     \bigg[ \sqrt{\det \Big( \delta^{\mu}_{~\nu}
     + \frac{\sigma}{m^2}G^{\mu}_{~\nu}\Big)}
     -1 - \frac{1}{2m^2l^2}\bigg]\,,
\end{equation}
where $G$ denotes the Einstein tensor,
$G^{\mu}_{~\nu}=R^{\mu}_{~\nu}-\half\delta^{\mu}_{~\nu}R$, and
$\eta$, $\sigma$ will be taken as $1$ or $-1$.

Expanding the square root expression, one obtains
\begin{equation}
 S = \frac{\eta}{2\kappa^2}\int d^3x \sqrt{-g}
     \bigg[\sigma  R+ \frac{2}{l^2} + \frac{1}{m^2}K
     - \frac{\sigma}{96m^4} K' +\CO(R^4)\bigg]\,.
\end{equation}
Note that this covers various cases by taking suitable signs of
$\eta$ and $\sigma$. Up to $R^3$ terms, the BI action becomes the
same form of the previous extended new massive gravity by choosing
$\mu^4 = 8m^4$ and $\xi =-\sigma$.  However, note that the sign of
the $R^3$ term is fixed in this case in terms of the sign of
Einstein-Hilbert term.

Equations of motion for the BI-NMG with $\sigma^2=1$ are given by
\begin{eqnarray}
 0 &=& \sqrt{\det \CA} \bigg[ 2\CB_{~(\mu}^{\alpha}R_{\nu)\alpha}
       -  \CB R_{\mu\nu} \bigg] - 2\sigma m^2 g_{\mu\nu}\Big( \sqrt{\det \CA}
       -   1-\frac{1}{2m^2l^2}\Big)  \nn \\
   &&  +   g_{\mu\nu} \bigg[\nabla_{\alpha}\nabla_{\beta}
       \Big(\sqrt{\det \CA}~\CB^{\alpha\beta}\Big)
       - \nabla^2\Big(\sqrt{\det \CA}~\CB\Big)\bigg]  \\
   &&  +  \nabla_{\mu}\nabla_{\nu}\Big(\sqrt{\det \CA}~\CB\Big)
       + \nabla^2\Big(\sqrt{\det \CA}~\CB_{\mu\nu}\Big)
       - \nabla^{\alpha}\nabla_{\mu}\Big(\sqrt{\det \CA}~\CB_{\nu\alpha}\Big)
       - \nabla^{\alpha}\nabla_{\nu}\Big(\sqrt{\det \CA}~\CB_{\mu\alpha}\Big) \,, \nn
\end{eqnarray}
where $\CA$ and $\CB$ are defined by
\[ \CA^{\mu}_{~\nu} \equiv \delta^{\mu}_{~\nu}
  + \frac{\sigma}{m^2}G^{\mu}_{~\nu}\,, \qquad
  \CB^{\mu}_{~\nu}  \equiv (\CA^{-1})^{\mu}_{~\nu}\,, \qquad
  \CB \equiv \CB^{\mu}_{~\mu}\,.
\]

%%%%%%%%%%%%%%%%%%%%%%%%%%%%%%%%%%%%%%%%%%%%%%%%%%%%%%%%%%%%%%%%%%%%%
\section{Black Hole Solutions}
%%%%%%%%%%%%%%%%%%%%%%%%%%%%%%%%%%%%%%%%%%%%%%%%%%%%%%%%%%%%%%%%%%%%%

There are various types of black holes in the above mentioned
gravity theories. It has already been known that BTZ black holes are
still solutions of any higher curvature  gravity theories in three
dimensions. Recently, it was shown that there are new type black
holes~\cite{Bergshoeff:2009aq} and warped $AdS$ black holes in
NMG~\cite{Clement:2009gq}. In this section, we will verify
explicitly the same form of the metrics are still solutions in the
extended NMG case. Our  main results are the verification of the
existence of such solutions and central charge expression of the
hypothetical dual CFTs.

Since we are dealing with higher curvature theories, the basic tool
in obtaining black hole entropy, which we will use, is the so-called
Wald formula, pioneered
in~\cite{Wald:1990sym}$\sim$\cite{Wald:1993nt}. This formula was
derived by interpreting  the black hole entropy as the Noether
charge on the horizon.  Using simplified version of this
formula~\cite{Jacobson:1993vj}\cite{Myers:1995}\cite{Saida:1999ec},
we will present our results for the entropy of various black holes.
By writing the black hole entropy in terms of the black hole
temperature and interpreting those quantities in the CFT side
according to the usual prescription of $AdS_3/CFT_2$ correspondence
for the parity symmetric case $c=c_L=c_R$ as
\begin{equation}
 S_{BH} = \frac{\pi^2 L}{3}(c_LT_L + c_RT_R)
        = \frac{\pi^2 L}{3}c(T_L + T_R) \,,
\end{equation}
we obtain the central charges of the dual CFTs.

According to the $AdS/CFT$ correspondence, the central charges
should be written in terms of the parameters in the dual gravity
Lagrangian but not of those in black holes, which means that the
central charge represents the property of theory but not of a
specific state. In other words,  when there are different black
holes with the same $AdS$ asymptotics, one can expect that the
central charges obtained through each black hole system should be
the same. In the $R^3$ extended NMG case, there are new type black
holes of the same asymptotic  with BTZ ones, which are,  henceforth,
expected to lead to the same central charges. In the following, we
will show that there are new type black holes in the $R^3$ extended
NMG, like the NMG case, when parameters in the Lagrangian take
specific values.  Indeed,  when parameters are chosen accordingly,
the central charge of the dual CFT obtained from these new type
black holes becomes identical with the central charge given by BTZ
black holes.

%\cite{Clement:1994gr} \cite{Clement:2003}\cite{Clement:2007}

We use the dimensional reduction procedure of
\cite{Clement:2009gq}\cite{Clement:1994}\cite{Clement:1994gr}$\sim$\cite{Clement:2008}
to find solutions with stationary circular symmetry. This method
have already been used to get  black hole solutions in TMG and NMG
case. According to this method, we can take a metric ansatz with two
Killing vectors $\partial_t$ and $\partial_{\phi}$ as
\begin{eqnarray}\label{metric}
 ds^2 = \lambda_{ab}(\rho) dx^a dx^b
          + \zeta(\rho)^{-2} R(\rho)^{-2} d\rho^2 \,,
\end{eqnarray}
where $x^0 = t, x^1 = \phi$, $R(\rho)^2 = -\det\lambda$ and
$\zeta(\rho)$ is the scale factor for arbitrary reparametrizations
of the radial coordinate $\rho$. The special linear group $SL(2,R)$
in the two Killing vector space is locally isomorphic to the Lorentz
group $SO(2,1)$, which suggests the parametrization of the matrix
$\lambda$
\begin{eqnarray}
 \lambda = \left(
              \begin{array}{cc}
                T(\rho)+X(\rho) & Y(\rho)  \\
                Y(\rho) & T(\rho)-X(\rho)
              \end{array}
           \right) \,,
\end{eqnarray}
such that special linear transformations of $\lambda$ correspond to
Lorentz transformations of the vector $\vec{X}=(T,X,Y)$ and $R^2 =
\vec{X}^2= \eta_{ij} X^i X^j = -T^2+X^2+Y^2$ is the norm in the
Minkowski space. As usual, we represent the scalar and vector
product of two vectors $\vec{X}$ and $\vec{Y}$ in the following form
\begin{eqnarray}
 \vec{X}\cdot\vec{Y} = \eta_{ij}X^i Y^j, ~~~~
    (\vec{X} \wedge \vec{Y})^i = \eta^{ij}\epsilon_{jkl}X^k Y^l \,,
\end{eqnarray}
with $\epsilon_{012} = +1$ for the wedge product.

For the chosen metric ansatz, we can obtain the Ricci tensor
components as~\cite{Clement:2009gq}\cite{Clement:2007}
\begin{eqnarray}
 {\cal R}^a_{~b} = -\frac{\zeta}{2} \left( (\zeta R R')'\bf{1}
           + (\zeta \ell)' \frac{}{} \right)^{a}_{~b} , ~~~~
 {\cal R}^{\rho}_{~\rho} = -\zeta(\zeta R R')'
 + \frac{1}{2}\zeta^2(\vec{X}'^2) \,, \qquad  ' \equiv \frac{d}{d\rho}\,,
\end{eqnarray}
where $\ell$ represents the matrix defined in terms of the
components of  the vector $\vec{L} = (L^T, L^X, L^Y) \equiv
\vec{X}\wedge\vec{X}'$
\begin{eqnarray}
  \ell = \left(
           \begin{array}{cc}
             -L^Y  &  -L^T + L^X  \\
             L^T + L^X & L^Y
           \end{array}
         \right)\,.
\end{eqnarray}
In the Appendix we write down the details of this method in our
case. Various curvature scalars for the action in the case of
$R^3$-NMG are given by
\begin{eqnarray}
 {\cal R} &=& \sigma \zeta^2 \left\{ -2(R R')'+\frac{1}{2}(\vec{X}'^2) \right\}
              - 2\sigma\zeta \zeta' R R' \,, \\
 K &=& \zeta^4 \left\{ \frac{1}{2}(\vec{L}'^2) - \frac{1}{4}(R R')'(\vec{X}'^2)
       + \frac{5}{32}(\vec{X}'^2)^2 \right\}   \nonumber\\
   &&  + \zeta\zeta'\left\{ (\vec{L}\cdot\vec{L}')
       - \frac{1}{4}R R'(\vec{X}'^2) \right\}
       + \frac{1}{2}\zeta^2 \zeta'^2(\vec{L}^2)  \,,  \\
 K' &=& \zeta^6 \left\{ -\frac{3}{2}(R R')'(\vec{X}'^2)^2
        + \frac{9}{8}(\vec{X}'^2)^3 + 24(R R')' (\vec{L}'^2)
        - 18(\vec{X}'^2)(\vec{L}'^2) \right\}  \nonumber\\
    &&  + \zeta^5\zeta' \left\{ -\frac{3}{2}(R R')'(\vec{X}'^2)^2
        + 24(R R')(\vec{L}'^2) + 48(R R')'(\vec{L}\cdot\vec{L}')
        \right.  \nonumber\\
    &&  \left. - 36(\vec{X}'^2)(\vec{L}\cdot\vec{L}') \right\}
        + \zeta^4 \zeta'^2 \left\{ 24(R R')' (\vec{L}^2)
        - 18(\vec{X}'^2)(\vec{L}\cdot\vec{L}') \right\}   \nonumber\\
    &&  + 24 \zeta^3 \zeta'^3 (R R')(\vec{L}^2)   \,.
\end{eqnarray}
Using these expressions, one can reduce the $R^3$-NMG action to the
one-dimensional form given in the Appendix up to a surface term.
From the reduced action, we can derive the reduced equations of
motion. After variation of $\zeta$ we can fix the gauge $\zeta =
const$ using the reparametrization invariance of the metric
(\ref{metric}). The first form (\ref{action-1}) of the reduced
action in the Appendix gives us the equation of motion after
variation with respect  to $\vec{X}$ as
\begin{eqnarray}\label{EOM-NMG}
\delta A + \frac{1}{\zeta^2} \delta E + \frac{1}{\zeta^4}
    \delta H = 0 \,,
\end{eqnarray}
where $\delta A, \delta E$ and $\delta H$ are written down
explicitly in Eq.(\ref{delta-A})$\sim$ (\ref{delta-H}) in the
Appendix. It is convenient to use the second form (\ref{action-2})
of the reduced action for the variation with respect to $\zeta$,
which gives us the following Hamiltonian constraint :
\begin{eqnarray}
 {\cal H} \equiv
 (5A-B')+\frac{1}{\zeta^2}(3E-F')+\frac{1}{\zeta^4}(H-J')
 -\frac{2}{\ell^2}\frac{1}{\zeta^6}
 =0 \,.
\end{eqnarray}
In the BI-NMG case, we can obtain the equation of motion through the
variation of the action with respect to $\vec{X}$
\begin{eqnarray}
0=  4(\det{\cal A})^2(\vec{S}'' - \vec{Q}' + \vec{P})
      - 4(\det{\cal A})(\det{\cal A}')\vec{S}'
      - 2(\det{\cal A})(\det{\cal A}'')\vec{S}  \nonumber\\
      + 3(\det{\cal A}')^2 \vec{S}
      + 2(\det{\cal A})(\det{\cal A}') \vec{Q}
       \,,
\end{eqnarray}
and from the variation with respect to $\zeta$, we can also obtain
the Hamiltonian constraint,
\begin{eqnarray}
0 =  \frac{1}{2}\frac{(\det{\cal A}')}{\det{\cal A}}{\cal N}
     + {\cal M} - {\cal N}' - 2(\det{\cal A})
     + 2(\det{\cal A})^{1/2} \left( 1+\frac{1}{2m^2\ell^2} \right)\,,
\end{eqnarray}
where ${\CM}$, $\CN$ and $\det{\cal A}$  are written down explicitly
in Eq.(\ref{Meq}),(\ref{Neq}) and (\ref{det1}) in the Appendix.
 The variation of the reduced action in both $R^3$-NMG and
BI-NMG cases are presented in detail in the Appendix.
\\\\
In order to obtain exact solutions, we can take a vector $\vec{X}$
as follows :
\begin{eqnarray}\label{vector}
 \vec{X} = \vec{\alpha} \rho^2 + \vec{\beta} \rho + \vec{\gamma} \,,
\end{eqnarray}
where $\vec{\alpha}, \vec{\beta}$ and $\vec{\gamma}$ are linearly
independent constant vectors. Substituting this ansatz into equation
(\ref{EOM-NMG}), we have to impose the following constraints on
vectors $\vec{\alpha}$ and $\vec{\beta}$ in order to vanish the
higher than the third order components which is given by
\begin{eqnarray}
 \vec{\alpha}^2 = 0,~~~(\vec{\alpha}\cdot\vec{\beta}) = 0  \,.
\end{eqnarray}
These constraints induce two more constraints
\begin{eqnarray}
 \vec{\alpha}\wedge\vec{\beta} = b\vec{\alpha}, ~~~ \vec{\beta}^2 =
 b^2 \,,
\end{eqnarray}
with a real constant $b$. Using these conditions with
$(\vec{\alpha}\cdot\vec{\gamma}) \equiv -z$, we can obtain the
equation of motion and the Hamiltonian constraint in the $R^3$-NMG
case as
\begin{eqnarray}
\!\!\!\!\!\!\!\!\!\!\!\!    0\!\! &=&\!   \left[ \xi\left(
-\frac{33}{8}b^4 + 30b^2z + 24z^2 \right)
      - \frac{\mu^4}{m^2\zeta^2}\left( \frac{17}{4}b^2 + 2z \right)
      - 2\frac{\mu^4}{\zeta^4} \sigma \right] \vec{\alpha} \,,
      \label{EQ1-NMG}\\
\!\!\!\!\!\! \!\!\!\!\!\!  0\!\!&=&\! \frac{\xi}{12 \mu^4} \left\{
-\frac{3}{8}b^6
      - 216b^2 z^2 + 36 b^4 z - 384 z^3 \right\}
      + \frac{1}{\zeta^2 m^2} \left\{ 2z^2 - \frac{1}{32}b^4 + 3b^2z \right\}
      + \frac{\sigma}{\zeta^4}\frac{b^2}{2} - \frac{2}{\ell^2
      \zeta^6}  \,. \label{HC-1}  \nn\\
\end{eqnarray}
In the BI-NMG case, we obtain
\begin{eqnarray}
  \det{\cal A} = \left( 1+\sigma\frac{\zeta^2}{m^2}\frac{b^2}{4}
          \right)^2
          \left\{ 1+\sigma\frac{\zeta^2}{m^2}
          \left( \frac{b^2}{4}-2z \right) \right\}   \,,
\end{eqnarray}
and note that $\det{\cal A}' = \det{\cal A}'' = 0$ under the above
conditions. The equation of motion and the Hamiltonian constraint
are given by
\begin{eqnarray}
 && 0 = \left[\sigma \frac{\zeta^2}{m^2} + \frac{5}{2}\frac{\zeta^4}{m^4}b^2
        + \sigma \frac{9}{16}\frac{\zeta^6}{m^6}b^4 \right]
        \vec{\alpha} \,,  \label{EQ2-BI}\\
 && 0 = 1+\sigma\frac{\zeta^2}{m^2}
        \left(\frac{1}{4}b^2-z\right)
        + \frac{\zeta^4}{m^4} \left( \frac{3}{4}b^2z \right)
        - \left( 1+\frac{1}{2m^2\ell^2} \right)
        \left[ 1+\sigma\frac{\zeta^2}{m^2}
        \left( \frac{1}{4}b^2-2z \right) \right]^{1/2}  \,.
        \label{HC-2} \nn\\
\end{eqnarray}
 As is given in the following, the case of $\vec{\alpha} = 0$
 in the Eq.s~(\ref{EQ1-NMG}) and (\ref{EQ2-BI}) leads to  BTZ
black holes, whereas the other condition of $\vec{\alpha} \neq 0$
leads to warped black hole solutions.

This section is organized as follows. BTZ black holes are presented
in subsection 3.1 with some explanation of generic features. In
subsection 3.2, we present warped black hole solutions. Finally, in
subsection 3.3, we present new type black holes and propose some
straightforward  extension of central charge function formalism for
applying it to new type black holes. %One of our main results are
%presented in subsection 3.3.

%%%%%%%%%%%%%%%%%%%%%%%%%%%%%%%%%%%%%%%%%%%%%%%%%%%%%%%%%%%%%%%%%%%%%
\subsection{BTZ black holes}
%%%%%%%%%%%%%%%%%%%%%%%%%%%%%%%%%%%%%%%%%%%%%%%%%%%%%%%%%%%%%%%%%%%%%

If we consider the condition $\vec{\alpha}=0$, then the vector
ansatz (\ref{vector}) reduces to
\begin{eqnarray}
 \vec{X} = \vec{\beta} \rho + \vec{\gamma} \,.
\end{eqnarray}
Taking two vectors $\vec{\beta}$ and $\vec{\gamma}$ to be
$\vec{\beta} = [ -(1-\ell^2)/\ell^2, -(1+\ell^2)/\ell^2, 0 ]$ and
$\vec{\gamma} = [ \tilde{m}(1+\ell^2)/4, \tilde{m}(1-\ell^2)/4, -j/2
]$,
then we find that the metric ansatz (\ref{metric}) becomes the BTZ
black hole solution\cite{BTZ:92}\cite{BTZ:1993}
\begin{eqnarray}
 ds^2 = \left( -\frac{2}{\ell^2}\rho + \frac{\tilde{m}}{2} \right) dt^2
          - j dt d\phi
          + \left( 2\rho + \frac{\tilde{m}\ell^2}{2} \right) d\phi^2
          + \frac{d\rho^2}{(\frac{4}{\ell^2}\rho^2
          - \frac{\tilde{m}^2\ell^2-j^2}{4})}  \,,
\end{eqnarray}
for $\zeta = 1$ and $b^2 = 4/\ell^2$\cite{Clement:2009gq}. If we
change the coordinate $\rho = r^2/2 - \tilde{m}\ell^2/4$, then we
can transform the $(t, \rho, \phi)$ coordinate system into the
$(t,r,\phi)$ one. We can also reparametrize $r \rightarrow \ell r,~t
\rightarrow -Lt$ and $\phi \rightarrow L\phi/\ell$ in order for $t$
and $r$ to be dimensionless. With $\zeta = \ell/L$ and $b^2 =
4/\ell^2$ we can see that BTZ black holes are solutions of the above
equations of motion, of which metric is given by
\begin{equation}\label{BTZ}
 ds^2 = L^2\bigg[ -\frac{(r^2-r^2_+)(r^2-r^2_-)}{r^2}dt^2
        +\frac{r^2}{(r^2-r^2_+)(r^2-r^2_-)}dr^2  + r^2\Big(d\phi
        + \frac{r_+r_-}{r^2}dt\Big)^2\bigg]\,,
\end{equation}
where  $L$ is the  $AdS$ length.  $L$ should  always be positive and
it  is natural to write physical quantities in terms of  this $AdS$
length, $L$.

The Hawking temperature of the black hole can be obtained by
dividing $2\pi $ of the surface gravity $\kappa$, which is, for the
above ADM form of the metric, given by
\[
 \kappa =  \Big(-\frac{1}{2} \nabla_{\mu}\xi_{\nu}\nabla^{\mu}\xi^{\nu} \Big)^{1/2}\bigg|_{r=r_+}
 = \frac{1}{L}\frac{\p_r N}{\sqrt{g_{rr}}}\bigg|_{r=r_+}\,, \]
where $\xi$ is a null Killing vector at the Horizon  with the
normalization as  $\xi^2 \rightarrow -r^2$ for $r\rightarrow
\infty$. This gives us
 \begin{equation}
  T_H =  \frac{r_+}{2\pi L}\Big(1-\frac{r^2_-}{r^2_+}\Big)\,.
\end{equation}
Another way to obtain the Hawking temperature is to demand the
regularity of the Euclideanized form of the black hole metric with
an appropriate  choice of time coordinate scale. The regularity
condition in this case leads to the periodicity of the Euclidean
time and then it can be interpreted as the temperature. Using this
approach, one obtains the same result with the above one in our
case. The angular velocity at the horizon can be defined as
\begin{equation}
 \Omega_H = \frac{1}{L} N^{\phi}(r_+) = \frac{1}{L}\frac{r_-}{r_+}\,.
\end{equation}
For BTZ black holes, the left and the right temperatures are given
by~\cite{Maldacena:1998}
\[
 T_L = \frac{r_+ +r_-}{2\pi L} \,, \qquad T_R = \frac{r_+ -r_-}{2\pi L} \,.
\]

For higher curvature gravity theories, there are some difficulties
to define conserved charges and we need new ingredients  to define
various physical quantities like mass and angular
momentum~\cite{Deser:2007vs}. Strictly speaking, it is required to
define various physical quantities in concrete ways and verify the
first law of black holes systematically. However, we will bypass
these by assuming the validity of the first law of black holes in
these higher curvature gravity theories. Then, the mass and angular
momentum of these black holes may be read by integrating the first
law of black hole thermodynamics
\[
 dM = T_HdS_{BH} +  \Omega_H dJ\,.
\]

Note that  the left and the right energies can also be defined as
\[ E_L \equiv \frac{\pi^2L}{6}c_LT^2_L\,, \qquad E_R \equiv \frac{\pi^2L}{6}c_RT^2_R\,,
\]
which are related to mass and angular momentum in the BTZ case as
\begin{equation}
M = E_L +E_R\,, \qquad J = L(E_L-E_R)\,.
\end{equation}
For the parity symmetric case, one can see, through the above
relations, that the mass and angular momentum are proportional to
the central charge $c=c_L=c_R$  of dual CFT, and check in both cases
that the integral Smarr relation holds as
\[
  M = \frac{1}{2}T_HS_{BH} + \Omega_HJ\,.
\]

\noindent\underline{$R^3$-NMG case}\\
The condition $\vec{\alpha} = 0$ gives the BTZ black hole solution
(\ref{BTZ}), then this solution has to satisfy the condition
(\ref{HC-1}). With $b^2 = 4/l^2$ and $\zeta = l/L$, $L^2$ should
satisfy
\begin{equation}
  \sigma -\frac{L^2}{l^2} - \frac{1}{4m^2 L^2}- \frac{\xi}{\mu^4 L^4} =0\,. \label{Eqa1}
\end{equation}
The entropy is
\begin{equation}
 S_{BH} = \frac{A_H}{4G}\eta\bigg(\sigma + \frac{1}{2m^2L^2}
          + \frac{\xi}{\mu^4L^4}\bigg)\,, \qquad A_H \equiv 2\pi Lr_+\,,
\end{equation}
where $L^2$ is given by the solution of Eq.~(\ref{Eqa1}). The mass
and angular momentum are given by
\begin{equation}
 M = \frac{r_+^2+r_-^2}{8G}\eta\bigg(\sigma + \frac{1}{2m^2L^2}
     + \frac{\xi}{\mu^4L^4}\bigg)\,, \qquad
 J = \frac{Lr_+r_-}{4G}\eta\bigg(\sigma + \frac{1}{2m^2L^2}
     + \frac{\xi}{\mu^4L^4}\bigg)\,.
\end{equation}
{}From Cardy formula, one  obtains the central charge  for of the
dual 2d CFT as
\begin{equation}
 c = \frac{3L}{2G}\eta\bigg[\sigma + \frac{1}{2m^2L^2}
     + \frac{\xi}{\mu^4L^4}\bigg]\,. \label{R3Cen}
\end{equation}
Note that the above mass and angular momentum are proportional to
the central charge, indeed.

For a comparison, let us consider the central charge of the dual CFT
obtained from the NMG $AdS$ black hole solution written in the same
form  with~(\ref{R3Cen}). In our convention\footnote{This is
consistent with results
in~\cite{Liu:0903nnm}$\sim$\cite{Bergshoeff:2009aq}.},  it is given
by
\[ c = \frac{3L}{2G}\eta\Big[\sigma+\frac{1}{2m^2L^2}\Big]\,, \]
where
\[
L^2 \equiv \frac{l^2}{2}\bigg[\sigma +
\sqrt{1-\frac{1}{m^2l^2}}\bigg]\,.\]
One can see that  $\eta=\sigma=1$  is the sign choice
in~\cite{Bergshoeff:2009aq} and $\eta=\sigma=-1$ is the choice
in~\cite{Clement:2009gq}.

%%%%%%%%%%%%%%%%%%%%%%%%%%%
\noindent\underline{BI-NMG case}\\
%%%%%%%%%%%%%%%%%%%%%%%%%%%
In this case we have to consider the constraint (\ref{HC-2}) with
$b^2 = 4/l^2$ and $\zeta = l/L$, then $L^2$ should satisfy
\[
\frac{1}{L^2} = \frac{\sigma}{l^2} \Big(1+\frac{1}{4m^2l^2}\Big)\,,
\]
where  $1/l^2$ is cosmological constant and may be negative, whereas
$AdS$ length, $L$, should be always positive.   One of interesting
points of the higher curvature theories is that there are
asymptotically $AdS$ solutions even though positive cosmological
constants.

The entropy in this case is given by
\begin{eqnarray}
 S_{BH} &=& -\eta\sigma\frac{A_H}{4G}\sqrt{\det \CA}\,
            \Big(\CB^{tt} (g^{tt})^{-1} +\CB^{rr}(g^{rr})^{-1}
            -\CB\Big)\Big|_{r=r_+}  \nn \\
    &=&  \eta\sigma\frac{A_H}{4G}\sqrt{1+\sigma\frac{1}{m^2L^2}}
         =\eta\sigma\frac{\pi  L r_+  }{2G}\sqrt{1+\sigma\frac{1}{m^2L^2}} \,.
\end{eqnarray}
By substituting $L$ in terms of $l$ and $m$, the entropy may be
written in terms of parameters in the Lagrangian with $r_+$, for
instance $\eta=\sigma=1$, as
\[
  S_{BH}=\frac{\pi  l  r_+ }{2G} \Big(1+ \frac{1}{2m^2l^2}\Big)
        \Big(1+\frac{1}{4m^2l^2}\Big)^{-1/2}\,.
\]
The mass of these black holes are given by
\begin{equation}
 M = \eta\sigma\frac{r_+^2+r_-^2}{8G}\sqrt{1+\sigma\frac{1}{m^2L^2}}\,, \qquad
 J = \eta\sigma\frac{Lr_+r_-}{4G}\sqrt{1+\sigma\frac{1}{m^2L^2}}\,.
\end{equation}
Noting that the Born-Infeld Lagrangian is also parity-symmetric, one
can obtain the central charge $c=c_L=c_R$ of the dual 2d CFT as
\begin{equation}
 c = \eta\sigma\frac{3L}{2G}\sqrt{1+ \sigma\frac{1}{m^2L^2}}  \,.
\end{equation}
Note that the positive central charge is allowed only when
$\eta\sigma=1$, which is the standard sign choice of the
Einstein-Hilbert term.

All the above central charges for BTZ black holes can also be
obtained by the central charge function formalism, which may be
thought as the compact summary of the above procedure:
\begin{equation}
 c \equiv \frac{L}{2G}g_{\mu\nu}\frac{\p \CL}{\p R_{\mu\nu}}\,.
\end{equation}
For instance, in the BI case
\[
  c = \eta\sigma \frac{L}{2G} \sqrt{\det \CA}\,\CB\,,
\]
where
\[
 \sqrt{\det \CA} = \Big(1+\sigma\frac{1}{m^2L^2}\Big)^{3/2}\,, \qquad
 \CB = 3\Big(1+\sigma\frac{1}{m^2L^2}\Big)^{-1}\,.
\]

%%%%%%%%%%%%%%%%%%%%%%%%%%%%%%%%%%%%%%%%%%%%%%%%%%%%%%%%%%%%%%%%%%%%%
\subsection{Warped AdS Solutions}
 %%%%%%%%%%%%%%%%%%%%%%%%%%%%%%%%%%%%%%%%%%%%%%%%%%%%%%%%%%%%%%%%%%%%

Setting the condition $\vec{\beta}^2 = b^2 =1$,
$(\vec{\beta}\cdot\vec{\gamma}) = 0$, $\vec{\gamma}^2 = -\beta^2
\rho_0^2$ and $(1-2z)=\beta^2$, then we can get $R^2 = (1-2z)\rho^2
+ \vec{\gamma}^2 = \beta^2(\rho^2-\rho_0^2)$. In addition to this
condition we can choose vectors as follows~\cite{Clement:2009gq}
\begin{eqnarray}
 && \vec{\alpha} = (1/2, -1/2, 0),~~~~~
    \vec{\beta} = (\omega, -\omega, -1),~~~ \nn\\
 && \vec{\gamma} = (z+u, z-u, -2\omega z)~~~~~
    (u = \beta^2\rho_{0}^2/4z + \omega^2 z)  \,,
\end{eqnarray}
which lead to the metric form
\begin{eqnarray}
 ds^2 = -\frac{\beta^2(\rho^2-\rho_0^2)}{\Delta^2}dt^2
        + \frac{d\rho^2}{\zeta^2\beta^2(\rho^2 - \rho_0^2)}
        + \Delta^2 \left( d\phi - \frac{\rho
        + (1-\beta^2)\omega}{\Delta^2}dt \right)^2  \,,
\end{eqnarray}
where $\Delta^2$ is defined by
\begin{eqnarray}
 \Delta^2 = \rho^2 + 2\omega \rho + 2u
          = \rho^2 + 2\omega \rho + (1-\beta^2)\omega^2
          + \frac{\beta^2 \rho_0^2}{1-\beta^2} \,.
\end{eqnarray}
If we take $\zeta^2 = 3/(4\beta^2-1)\cdot4/L^2$ and reparametrize $t
\rightarrow t/(\zeta\beta^2)$ and $\phi \rightarrow \phi/\zeta$,
then we can obtain warped AdS black hole solutions as follows:
\begin{equation}\label{Warped-BH}
 ds^2 = \frac{4\beta^2-1}{12\beta^2}L^2 \bigg[ -\frac{\rho^2-\rho^2_0}{\Delta^2}dt^2
        + \frac{d\rho^2}{\rho^2-\rho^2_0} +\beta^2\Delta^2\Big(d\phi
        - \frac{\rho+(1-\beta^2)\omega}{\beta^2\Delta^2}dt\Big)^2\bigg]\,,
\end{equation}
where we have chosen the overall coefficient such that $L$
corresponds to the Ricci scalar curvature  as $
 R =  - \frac{6}{L^2} \,.
$
To see this, one may note that the asymptotic region  is given by
taking $\rho\rightarrow \infty$ as
\begin{equation}
 ds^2  \simeq \frac{4\beta^2-1}{12\beta^2}L^2\bigg[ - dt^2
      + \frac{d\rho^2}{\rho^2} +\beta^2\rho^2\Big(d\phi
      - \frac{1}{\beta^2}\frac{1}{\rho}dt\Big)^2\bigg]\,,
\end{equation}
which can be shown to be the warped $AdS_3$ in Poincar\'{e}
coordinates by a suitable change of variables.\footnote{The
coordinate transformations are not unique at the asymptotic region.
For coordinate transformation to the Schwarzschild type,
see~\cite{Anninos:2008fx}.} It is an interesting fact that the above
warped $AdS_3$ black hole metric has constant curvature invariants
of the same values with the warped $AdS_3$, which allows to obtain
it by the quotienting method from the warped $AdS_3$
space~\cite{Kraus:2005vz}\cite{Anninos:2008fx}.

By a suitable coordinate transformation  with parameters relations
\[ \rho_0 = \frac{1}{2}(r_+ -r_-)\,, \qquad (1-\beta^2)\omega
          = \frac{1}{2}\Big(r_+ +r_- -2\beta\sqrt{r_+r_-}\Big)\,, \qquad
 \beta^2 = \frac{\nu^2+3}{4\nu^2}\,,
\]
the above metric can be written as~\cite{Anninos:2008fx}
\begin{equation}
 ds^2 =  L^2 \bigg[-N(r)^2dt^2 +R(r)^2(d\theta +N^{\theta}(r)dt)^2
      + \frac{ dr^2}{4R(r)^2N(r)^2} \bigg]\,,
\end{equation}
where
\begin{eqnarray}
 N(r)^2 &=& \frac{(\nu^2+3)(r-r_+)(r-r_-)}{4R(r)^2}\,,\nn \\
N^{\theta}(r) &=& \frac{2\nu r-\sqrt{r_+ r_- (\nu^2+3)}}{2R(r)^2} \,, \nn \\
R(r)^2 &=& \frac{r}{4} \left( 3(\nu^2-1)r + (\nu^2+3)(r_+ + r_-) -
4\nu\sqrt{r_+ r_-(\nu^2+3)} \right) \,. \nn
\end{eqnarray}
We will use this form of the metric for the entropy, Hawking
temperature, etc in the following.

The black hole temperature and the angular velocity are given by
\begin{equation}
 T_H =  \frac{\beta^2}{2\pi L}\sqrt{\frac{3}{4\beta^2-1}}\,
        \left(\frac{  r_+ - r_-}{r_+ - \beta\sqrt{r_+r_-}} \right)\,, \qquad
 \Omega_{H}  = \frac{1}{L}\sqrt{\frac{4\beta^2-1}{3}}\,
        \left( \frac{1}{r_+-\beta\sqrt{r_+r_-}} \right)\,.
\end{equation}
The left and the right temperatures of these warped $AdS$ black
holes are given by~\cite{Anninos:2008fx}
\begin{equation}
 T_L =  \frac{r_+ +r_- -2\beta\sqrt{r_+r_-}}{2\pi L}
        \left(\frac{3\beta^2}{4\beta^2-1} \right) \,, \qquad
 T_R =  \frac{r_+ -r_-}{2\pi L}  \left( \frac{3\beta^2}{4\beta^2-1} \right)\,,
\end{equation}
which are related to the above Hawking temperature and angular
velocity as
\[
  \frac{1}{T_H} = \pi L \sqrt{\frac{4\beta^2-1}{3\beta^4}}
        \left(\frac{T_L + T_R}{T_R} \right)\,, \qquad
  \frac{\Omega_H}{T_{H}} = \frac{1}{T_R L}\,.
\]
It was suggested in~\cite{Anninos:2008fx} that more useful charges
in the warped case are the left and the right moving energies which
are defined by
\[ E_L \equiv \frac{\pi^2 L}{6}c_LT^2_L\,, \qquad
   E_R \equiv \frac{\pi^2 L}{6}c_RT^2_R\,. \]
By construction, these charges satisfy
\begin{equation}
  \frac{\p S_{BH}}{\p E_L } = \frac{1}{T_L}\,, \qquad
  \frac{\p S_{BH}}{\p E_R } = \frac{1}{T_R}\,. \label{WarpedFirstLaw}
\end{equation}
The mass and angular momentum can be related to these charges
as\footnote{We have modified slightly the relations given
in~\cite{Anninos:2008fx} in order to apply these to our case. These
modified ones also lead to the correct results in TMG case.}
\begin{equation}
  M  = \sqrt{\frac{3\beta^4}{4\beta^2-1}}\sqrt{\frac{2c_L E_L}{3L}} \,, \qquad
  J = L (E_L-E_R)\,.
\end{equation}
One can check that $(M, J)$ defined in the above satisfy, through
the formula~(\ref{WarpedFirstLaw}),  the first law of black holes,
\[ dM = T_HdS_{BH}+\Omega_H dJ\,. \]
By using the left and the right temperatures in terms of black hole
parameters $r_+, r_-$ and recalling $c_L=c_R =c$ in our case, one
may write the mass and the angular momentum in terms of the central
charges as follows:
\begin{eqnarray}
 M &=& \frac{\beta}{6L}\left(\frac{3\beta^2}{4\beta^2-1}\right)^{3/2}~
       \Big[r_+ + r_- -2\beta\sqrt{r_+r_-}\Big] \, c \,, \\
 J &=& \frac{1}{24}\left(\frac{3\beta^2}{4\beta^2-1}\right)^{2}
       \Big[ (r_++r_--2\beta\sqrt{r_+r_-})^2 - (r_+-r_-)^2\Big]\, c\,,
\end{eqnarray}
where $c$ denotes the central charge of dual CFT. One may also write
$(M,J)$ in terms of the black entropy since the central charges are
proportional to the entropy in the present case.

Using  black hole entropy, $S_{BH}$, given in the following, one can
show that the above formula satisfies Smarr-like relation
\[ M = T_H S_{BH} +2\Omega_HJ\,,\]
and verify the differential form of the first law of black holes
explicitly. We will not present the mass and angular momentum
explicitly in  each of the extended NMG's , since it is
straightforward to obtain those for the given central charges which
are given in the following.

%%%%%%%%%%%%%%%%%%%%%%%%%%%%%%%%%%%%%%%%%%%%
\noindent\underline{$R^3$-NMG case}\\
%%%%%%%%%%%%%%%%%%%%%%%%%%%%%%%%%%%%%%%%%%%
Substituting the values $b^2 = 1, z=(1-\beta^2)/2$ and $\zeta^2 =
3/(4\beta^2-1)\cdot4/L^2$ into two equations (\ref{EQ1-NMG}) and
(\ref{HC-1}),then we can see that $L$ and $\beta$ should satisfy two
equations
\begin{eqnarray}
0 &=& \sigma - \frac{3(4\beta^2-21)}{2(4\beta^2-1)}\frac{1}{m^2 L^2}
      -\xi \frac{27(4\beta^2-3)(4\beta^2-15)}{(4\beta^4-1)^2}\frac{1}{\mu^4L^4}\,,  ~\nn \\
0 &=& \sigma -\frac{4\beta^2-1}{3} \frac{L^2}{l^2}
      + \frac{3(16\beta^4-80\beta^2+63)}{4\beta^2-1}\frac{1}{4m^2L^2}
      + \frac{9(4\beta^2-3)(32\beta^4-108\beta^2+75)}{(4\beta^4-1)^2}
      \frac{\xi}{\mu^4 L^4}\,. ~\nn
\end{eqnarray}
Note that $L$ can be solved as
\[
 L^2 =  \frac{\sigma}{(4\beta^2-1) }\frac{3}{4m^2\mu^2}
        \bigg\{(4\beta^2-21)\mu^2\pm \sqrt{(4\beta^2-21)^2\mu^4
        +48\xi\sigma(4\beta^2-3)(4\beta^2-15)  m^4}\bigg\}\,.
\]
Using Wald's formula, one obtains
\begin{eqnarray}
 S_{BH}  &=& \frac{A_H}{4G}\eta
         \bigg[\sigma + \frac{3(5-4\beta^2)}{2(4\beta^2-1) m^2 L^2}
         - \xi\frac{9(3-4\beta^2)(13-12\beta^2)}{(4\beta^4-1)^2\mu^4L^4}\bigg]\,,  \\
  A_H &\equiv& 2\pi L R(r_+) = 2\pi L \sqrt{\frac{3}{4\beta^2-1}}
         \Big(r_+ -\beta \sqrt{r_+r_-}\Big)\,, \nn
\end{eqnarray}
where  $\beta$ and $L^2$ are given by the solutions of the above
equation.

The central charges for the warped $AdS$ black holes can also be
obtained by the Cardy formula
\[ S = \frac{\pi^2 L}{3}(c_LT_L +c_R T_R) = \frac{\pi^2 L}{3}c(T_L +T_R)\,, \]
which leads to
\begin{equation}
 c = \frac{L}{2G}\frac{\sqrt{3(4\beta^2-1)}}{\beta^2} \eta
     \bigg[\sigma + \frac{3(5-4\beta^2)}{2(4\beta^2-1) m^2 L^2}
     - \xi\frac{9(3-4\beta^2)(13-12\beta^2)}{(4\beta^4-1)^2\mu^4L^4}\bigg]\,. \label{CenR3}
\end{equation}
The naive application of the central charge function formalism leads
to
\begin{equation}
 c = \frac{3L}{2G}\eta\bigg[ \sigma  + \frac{1}{2m^2L^2}
     -\xi \frac{3(4\beta^2-3)(4\beta^2-7)}{(4\beta^4-1)^2\mu^4L^4}\bigg]\,,
\end{equation}
which  gives the different answer in this case, which is not
unexpected since the central charge function formalism is developed
for the $AdS$ space not warped $AdS$ space.

One can see that the above result~(\ref{CenR3}) reproduces the new
massive gravity case if one set $\xi=0$
\begin{eqnarray}
\beta^2 &=& \frac{1}{4(1-m^2l^2)}\Big(21-9m^2l^2
            \pm 2\sqrt{3(7+5m^2l^2)m^2l^2}\Big)\,, \nn \\
\frac{1}{L^2} &=&  \frac{2(4\beta^2-1) \sigma m^2}{3(4\beta^2-21)}
            = \frac{2\sigma}{63l^2}\Big(-39m^2l^2
            \pm 10\sqrt{3(7+5m^2l^2)m^2l^2}\Big)\,,
\end{eqnarray}
which implies
\[
 S_{BH}  = \frac{A_H}{4G}\eta\bigg[\sigma
           + \frac{3(5-4\beta^2)}{2(4\beta^2-1) m^2 L^2}\bigg]
         = \frac{A_H}{4G}\frac{16\eta\sigma}{21-4\beta^2}
         = \frac{A_H}{4G}\eta\Big(\frac{4}{5}\sigma -\frac{6}{5m^2 L^2}\Big)\,,
\]
and
\[ c = \eta\sigma \frac{8L}{G} \frac{\sqrt{3(4\beta^2-1)}}{\beta^2(21-4\beta^2)} \,.\]
These results are consistent with the choice of $\eta =\sigma =-1$
in~\cite{Kim:2009jm}.

\noindent\underline{BI-NMG case}\\
Substituting the values $b^2 = 1, z=(1-\beta^2)/2$ and $\zeta^2 =
3/(4\beta^2-1)\cdot4/L^2$ into the equations (\ref{EQ2-BI}) and
(\ref{HC-2}), we obtain
\begin{eqnarray}
 L^2 &=&  -\frac{27\sigma}{(4 \beta^2-1) m^2}\,,  \label{Lsq}\\
 0 &=& 1 + \frac{3(2\beta^2-1)}{4\beta^2-1}\frac{\sigma}{m^2L^2}
       -\frac{54(\beta^2-1)}{(4\beta^2-1)^2}\frac{1}{m^4L^4}
       -\Big(1+\frac{1}{2m^2l^2} \Big)
       \Big[1+\frac{3(4\beta^2-3)}{4\beta^2-1}\frac{\sigma}{m^2L^2}\Big]^{1/2}
       \,,\nn
       \label{beta}~~~~~\\
\end{eqnarray}
where we may note that there are warped solutions only for
$\sigma=-1$. \\
Using Wald's formula, one obtains ($ A_H \equiv 2\pi L R(r_+) $)
\begin{eqnarray}
 S_{BH} &=& -\eta\sigma\frac{A_H}{4G}\sqrt{\det \CA}\,
            \Big(\CB^{tt} (g^{tt})^{-1} +\CB^{rr}(g^{rr})^{-1}
            -\CB\Big)\Big|_{\rho=\rho_0} \,,     \nn \\
        &=& \eta\sigma\frac{A_H}{4G}
            \bigg(1+\frac{3\sigma}{(4\beta^2-1)m^2L^2}\bigg)
            \bigg[1+\sigma\frac{3(4\beta^2-3)}{(4\beta^2-1) m^2
            L^2}\bigg]^{-1/2}\,.
\end{eqnarray}
Using the Cardy formula, we also get the central charges,
\begin{equation}
c  =  \eta\sigma \frac{L}{2G}\frac{\sqrt{3(4\beta^2-1)}}{\beta^2}
      \bigg(1+\frac{3\sigma}{(4\beta^2-1)m^2L^2}\bigg)
      \bigg[1+\sigma\frac{3(4\beta^2-3)}{(4\beta^2-1) m^2 L^2}\bigg]^{-1/2}\,,
\end{equation}
 where $L$ and $\beta$ are given by the solution of Eqs~(\ref{Lsq}) and (\ref{beta}).

 These may be written as
\begin{eqnarray}
 S_{BH} &=& \eta\sigma \frac{A_H}{4G} \frac{4}{3\sqrt{3-\beta^2}}
            = \eta\sigma \frac{A_H}{4G} \frac{8}{3}
            \Big(11+\frac{27\sigma}{m^2L^2}\Big)^{-1/2}\,, \nn \\
 c &=& \eta\sigma \frac{2L}{G} \sqrt{\frac{(4\beta^2-1)}{ 3(3-\beta^2)}}
            \frac{1}{\beta^2}\,. \nn
\end{eqnarray}
The central charge function formalism leads to central charge of the
dual CFT as
\begin{equation}
 c = \eta\sigma \frac{L}{2G} \sqrt{\det \CA}\,
 \CB=  \frac{2L}{3G} \eta \frac{(\beta^2-4)}{\sqrt{3-\beta^2}} \,,
\end{equation}
which is different from the above one.

%%%%%%%%%%%%%%%%%%%%%%%%%%%%%%%%%%%%%%%%%%%%%%%%%%%%%%%%%%%%%%%%%%%%%%
\subsection{New Type Black Holes}
%%%%%%%%%%%%%%%%%%%%%%%%%%%%%%%%%%%%%%%%%%%%%%%%%%%%%%%%%%%%%%%%%%%%%%
There are new type black hole solutions, which exist for a specific
values of parameters in the Lagrangian in the $R^3$-NMG
case.\footnote{This is already pointed out in~\cite{Sinha:2010ai}.}
One can check these type black holes solutions exist only in the
$R^3$-NMG case. These black holes are shown to exist  in the NMG
case, but their properties are not investigated in detail even in
the NMG case. (However, see~\cite{Giribet:2009qz} for some study of
their properties and extension to the rotating new type black holes
in the NMG case.) Its metric is given by
\begin{equation}
  ds^2 = L^2 \bigg[ -(r^2+br+c)dt^2 +\frac{dr^2}{r^2+br+c} + r^2 d\phi^2\bigg]\,.
\end{equation}
When   $b$ is zero, this reduces the  BTZ case of $r_-=0$.

 To satisfy the EOM's, $l^2$ should satisfy the following equations
\begin{equation}
  l^2 = \frac{(12 \sigma \xi  m^4  +\mu^4) L^2 +6\xi m^2 }{m^2(8\xi m^2 + \mu^4L^2)}\,,
\end{equation}
and $L^2$ is given by
\begin{equation}
 L^2 = \frac{\sigma}{4m^2\mu^2}
       \Big(\mu^2\pm \sqrt{\mu^4+48\sigma \xi m^4}\Big)\,. \label{NBHLsq}
\end{equation}
Note that in the NMG limit (or $\xi\rightarrow 0$) these conditions
become
\[
 l^2= \frac{1}{m^2}\,, \qquad L^2 = \frac{1}{2m^2}\,, \qquad \sigma=1\,.
\]

For the $b\neq 0$ case, the black hole horizon is at $r_{\pm}=(-b\pm
\sqrt{b^2-4c})/2$ and the entropy and the temperature of the black
holes are given by
\begin{eqnarray}
 S_{BH} &=& \frac{\pi L}{2G}\frac{\sqrt{b^2-4c}}{3}\eta\bigg[2\sigma
            + \frac{1}{2m^2L^2}\bigg]\,, \nn \\
            %  \frac{A_H}{4G}\eta\bigg[\sigma + \frac{1}{2m^2L^2}
            %  + \frac{\xi}{\mu^4L^4} - \bigg(\frac{1}{4m^2L^2}
            %  + \frac{\xi}{\mu^4L^4}\bigg)\frac{b}{c}\bigg(b+\sqrt{b^2-4c}\bigg)\bigg]\,,
            %  \qquad A_H \equiv 2\pi L r_+\,,\nn \\
 T_{H} &=& \frac{\sqrt{b^2-4c}}{4\pi L}=\frac{r_+-r_-}{4\pi L}\,,\nn
\end{eqnarray}
where we have used  the value of $L^2$ given in Eq.~(\ref{NBHLsq})
to simplify the expression for the entropy. For the above static
black hole solutions, one may use $AdS/CFT$ dictionary $S_{BH} =
\frac{2\pi^2 L}{3}c T_H$ to obtain the central charge as
\begin{equation}
 c= \frac{L}{G}\eta \bigg[ 2\sigma + \frac{1}{2m^2L^2}\bigg]\,,
\end{equation}
where $L^2$ is given in the Eq.~(\ref{NBHLsq}). This result is
consistent with~\cite{Oliva:2009ip} in the limit of $\xi=0$ with
$\eta=\sigma=1$.  One can see that this central charge is identical
with~(\ref{R3Cen}) by noting that    the chosen value of $L^2$
in~(\ref{NBHLsq})  leads to
\[
  \frac{\xi}{\mu^4 L^4} = \frac{1}{3}\Big(\sigma - \frac{1}{2m^2L^2}\Big)\,.
\]

One way to define the mass of these black holes resorts to the
$AdS/CFT$ dictionary through the relation $M =E_L +E_R$.  In this
case, this relation becomes $M= \frac{\pi^2L}{3}cT^2_H$, which leads
to
\begin{equation}
 M = \frac{b^2-4c}{48G}\eta\Big[ 2\sigma + \frac{1}{2m^2L^2}\Big]\,,
\end{equation}
This mass formula satisfies the simple form of the first law of
black holes, $dM = T_HdS_{BH}$ and reduces to the correct value in
the BTZ limit, $b=0$. However, it is unclear how to apply the first
law of black holes in this case,  since the nature of parameter  $b$
is obscure.  In fact, there is an attempt to realize the first law
of black holes with the parameter $b$ and  to understand its meaning
as a new gravitational hair~\cite{Giribet:2009qz}.

For new type black holes, the central charge function formalism
leads to non-constant central charge. This fact is simply the
indication that the curvature invariants of new type black holes are
not constant. Therefore, we need some modification of the known
central charge function formalism in this case. However, by
recalling the fact that dual CFT resides in the boundary of bulk
$AdS$ space, it is very suggestive to define the central charge in
this case as
\begin{equation}
  c = \eta\sigma \frac{L}{2G} \sqrt{\det \CA}\,\CB\bigg|_{r\rightarrow \infty}\,,
\end{equation}
where $r$ is the radial coordinate and $r=\infty$ denotes  the
position of the boundary.  Because the value obtained in this way is
identical with the one by the Cardy formula, this formula is a
natural generalization of the usual central charge function formula.

%%%%%%%%%%%%%%%%%%%%%%%%%%%%%%%%%%%%%%%%%%%%%%%%%%%%%%%%%%%%%%%%%%
\section{Conclusion}
%%%%%%%%%%%%%%%%%%%%%%%%%%%%%%%%%%%%%%%%%%%%%%%%%%%%%%%%%%%%%%%%%
We have verified that BTZ, warped $AdS$ black holes and new type
black holes are solutions in the extended new massive gravity
theories, and investigated their properties in the view point of the
AdS/CFT correspondence.  \\
\indent Firstly, we have presented various physical quantities,
mass, angular momentum, Hawking temperature and entropy, of black
holes in all the cases. Secondly, we have obtained central charges
of hypothetical dual CFT using Cardy formula and central charge
function formalism. Our results of entropy and central charges
reduce to the known NMG cases  in the NMG limit. \\
\indent We have considered two version of extended NMG, $R^3$-NMG
and BI-NMG, which seem to be related intimately.  One may regard
BI-NMG as a natural extension of $R^3$-NMG to the case of an
infinite number of higher curvature  terms. However, it needs to be
clarified why the new type black holes exist only for the $R^3$-NMG
case and  whether the  BI-NMG can be derived by the same argument
for the extension of NMG  to $R^3$-NMG.

We have obtained central charges of the dual CFTs by assuming the
validity of Cardy formula, which is not self-transparent and may be
thought as conjectural. Therefore,  it will be interesting to obtain
the central charges in other ways and verify our results, for
example, by the approach
in~\cite{Brown-Henneaux:86}\cite{detournay:0808}$\sim$\cite{cvetkovic:0907}.
Mass and angular momentum are also derived through the $AdS/CFT$
dictionary. It is required to obtain these quantities as conserved
charges like
in~\cite{Deser:2007vs}.  \\
\indent We have verified that the central charge function formalism
leads to the same results with the Cardy formula for asymptotically
$AdS$ black holes including new type black holes. For the new type
black holes, we need a simple extension of central charge function
formalism to match the results. However, in warped $AdS$ black
holes,  central charge function formalism, which is developed mainly
for BTZ black holes,  leads to different results from Cardy formula.
It is very interesting to extend central charge function formalism
to the warped $AdS$ case.

There are various direction to pursue in the future. First, though
the gravitons are massive in the NMG case, it is not yet known that
is the case for the extended NMG case.  Therefore, it is necessary
to analyze the linearized theory to see the nature of gravitons, and
to see whether massive gravitons and positive mass black holes are
compatible, which was not the case in the NMG or TMG case. It will
also be interesting to study the possibility of string theory
embedding or the supersymmetric extension of extended NMG. \\
\indent Finally, new type black holes should be studied in more
detail to see their meaning. They seem to have new type
gravitational hair which is not yet understood completely. The fact
that BI-NMG doesn't allow the new type black holes may also be
addressed to see whether the existence of new type black holes is
artifact or leads to some important  lessons.

%%%%%%%%%%%%%%%%%%%%%%%%%%%%%%%%%%%%%%%%%%%%%%%%%%%%%%%%%%%%%%%%%
\section*{\center Acknowlegements}
%%%%%%%%%%%%%%%%%%%%%%%%%%%%%%%%%%%%%%%%%%%%%%%%%%%%%%%%%%%%%%%%%
S.H.Y would like to thank the string theory group at Yonsei
University, especially, Prof. Seungjoon Hyun for enlightening
discussion.  S.H.Y was supported by the National Research Foundation
of Korea (NRF) grant funded by the Korea government (MEST) with the
grant number 2009-0085995. This work of S.N and J.D.P was supported
by a grant from the Kyung Hee University in 2009(KHU-20100130). S.N
was also supported by the National Research Foundation of Korea(NRF)
grant funded by the Korean government(MEST)(No. 2009-0063068).

\appendix
  \begin{center}
    {\bf APPENDIX}
  \end{center}
  \renewcommand{\theequation}{A.\arabic{equation}}
  \setcounter{equation}{0}
In the NMG case, the reduced one-dimensional action derived from the
action (\ref{NMG}) is given by
\begin{eqnarray}
 S_1 &=& \int d\rho [ A\zeta^5 + B\zeta^4\zeta' + C\zeta^3\zeta'^2 + D\zeta\zeta'^3
                  + E\zeta^3 + F\zeta^2\zeta' + G\zeta\zeta'^2  \nonumber \\
   && ~~~ + H\zeta + J\zeta' + \frac{2}{\ell^2}\zeta^{-1} ]  \,, \label{action-1}  \\
   &=& \int d\rho \left[ \left( A-\frac{1}{5}B' \right)\zeta^5 + C\zeta^3\zeta'^2 + D\zeta^2\zeta'^3
                  + \left( E-\frac{1}{3}F' \right)\zeta^3 + G\zeta\zeta'^2 \right. \nonumber\\
   && ~~~ \left. + (H-J')\zeta + \frac{2}{\ell^2}\zeta^{-1}\right]
   \,,
   \label{action-2}
\end{eqnarray}
where
\begin{eqnarray*}
 A &=& \frac{\xi}{12 \mu^4} \left\{ -\frac{3}{8}(\vec{X}'^2)^3
           -\frac{3}{2}(\vec{X}'^2)^2 (\vec{X}\cdot\vec{X}'')
           + 6[(\vec{X}'^2) + 4(\vec{X}\cdot \vec{X}'')]
           (\vec{X} \wedge \vec{X}'')^2  \right\}  \,,\\
 B &=& \frac{\xi}{12 \mu^4} \left\{ -\frac{3}{2}(\vec{X}'^2)^2 (\vec{X}\cdot\vec{X}')
       + 24(\vec{X}\cdot\vec{X}')(\vec{X}\wedge\vec{X}'')^2
       \right.  \\
   &&  \left. + 12[(\vec{X}'^2) + 4(\vec{X}\cdot\vec{X}'')]
       (\vec{X}\wedge\vec{X}')\cdot(\vec{X}\wedge\vec{X}'')   \right\}    \,, \\
 E &=& \frac{1}{m^2} \left\{ \frac{1}{2}(\vec{X}\wedge\vec{X}'')^2
       -\frac{1}{4}(\vec{X}'^2)(\vec{X}\cdot\vec{X}'') -\frac{3}{32}(\vec{X}'^2)^2
           \right\} \,,\\
 F &=& \frac{1}{m^2} \left\{ (\vec{X}\wedge\vec{X}')\cdot(\vec{X}\wedge\vec{X}'')
       -\frac{1}{4}(\vec{X}'^2)(\vec{X}\cdot\vec{X}')
           \right\}   \,, \\
 H &=& - \sigma\left\{\frac{3}{2}(\vec{X}'^2)
       + 2(\vec{X}\cdot\vec{X}'')\right\}   \,,\\
 J &=& -2\sigma(\vec{X}\cdot\vec{X}')  \,,
\end{eqnarray*}
and $(\vec{A} \wedge \vec{B}) \cdot (\vec{C} \wedge \vec{D}) =
-(\vec{A}\cdot\vec{C})(\vec{B}\cdot\vec{D}) +
(\vec{A}\cdot\vec{D})(\vec{B}\cdot\vec{C})$. Other terms are not
needed to get equation of motion and Hamiltonian constraint.   \\
The variation relative to $\vec{X}$ for the first form of the action
(\ref{action-1}) gives the equation of motion as following
\begin{eqnarray}
 \delta A + \frac{1}{\zeta^2} \delta E + \frac{1}{\zeta^4}
    \delta H = 0 \,,
\end{eqnarray}
where
\begin{eqnarray}
 \delta A &=& \frac{\xi}{12\mu^4} \left\{ (12(\vec{X'^2})
              +48(\vec{X}\cdot\vec{X}''))[\vec{X}\wedge(\vec{X}''''\wedge\vec{X})]
              \right.    \nonumber\\
          &&  \left. + (24(\vec{X}'^2)+96(\vec{X}\cdot\vec{X}''))
              [ \vec{X}'\wedge(\vec{X}'''\wedge\vec{X})
              + \vec{X}\wedge(\vec{X}'''\wedge\vec{X}')
              + \vec{X}'\wedge(\vec{X}''\wedge\vec{X}') ]
              \right.  \nonumber\\
          &&  \left. + 6(\vec{X}'^2)[\vec{X}'''\wedge(\vec{X}\wedge\vec{X}')
              + \vec{X}''\wedge(\vec{X}\wedge\vec{X}'')]
              + 12(\vec{X}'\cdot\vec{X}'')[\vec{X}''\wedge(\vec{X}\wedge\vec{X}')]
              \right.  \nonumber\\
          &&  \left. + (144(\vec{X}'\cdot\vec{X}'')
              +96(\vec{X}\cdot\vec{X}'''))[\vec{X}\wedge(\vec{X}'''\wedge\vec{X})
              + \vec{X}'\wedge(\vec{X}''\wedge\vec{X})
              + \vec{X}\wedge(\vec{X}''\wedge\vec{X}') ] \right. \nonumber \\
          &&  \left. + (72(\vec{X}''^2)+120(\vec{X}'\cdot\vec{X}''')
              +48(\vec{X}\cdot\vec{X}''''))[\vec{X}\wedge(\vec{X}''\wedge\vec{X})] \right.  \nonumber\\
          &&  \left. -\frac{3}{4}(\vec{X}'^2)^2 \vec{X}'' + 36(\vec{X}\wedge\vec{X}'')^2 \vec{X}''
              + 3(\vec{X}'^2)(\vec{X}'\cdot\vec{X}'')\vec{X}'  \right.    \nonumber\\
          &&  \left. + 72(\vec{X}\wedge\vec{X}'')\cdot(\vec{X}'\wedge\vec{X}'')\vec{X}'
              + 72(\vec{X}\wedge\vec{X}'')\cdot(\vec{X}\wedge\vec{X}''')\vec{X}' \right.   \nonumber\\
          &&  \left. + 48(\vec{X}'\wedge\vec{X}'')^2\vec{X} + 48(\vec{X}\wedge\vec{X}''')^2\vec{X}
              + 48(\vec{X}\wedge\vec{X''}) \cdot(\vec{X}\wedge\vec{X}'''')\vec{X}  \right.  \nonumber\\
          &&  \left. + 96(\vec{X}'\wedge\vec{X}'')\cdot(\vec{X}\wedge\vec{X}''')\vec{X}
              + 96(\vec{X}\wedge\vec{X}'') \cdot(\vec{X}'\wedge\vec{X}''')\vec{X} \right\} \cdot
              \delta\vec{X} \,,
              \label{delta-A}\\
 \delta E &=& -\frac{1}{m^2} \left\{ \vec{X}\wedge(\vec{X}\wedge\vec{X}'''')
              + \frac{5}{2}\vec{X}\wedge(\vec{X}'\wedge\vec{X}''')
              + \frac{3}{2}\vec{X}'\wedge(\vec{X}\wedge\vec{X}''') \right. \nonumber\\
          &&  \left. + \frac{9}{4}\vec{X}'\wedge(\vec{X}'\wedge\vec{X}'')
              - \frac{1}{2}\vec{X}''\wedge(\vec{X}\wedge\vec{X}'')
              - \frac{1}{8}(\vec{X}'^2)\vec{X}'' \right\}\cdot
              \delta\vec{X} \,, \label{delta-E} \\
 \delta H &=& -\sigma\vec{X}''\cdot \delta\vec{X}  \label{delta-H} \,.
\end{eqnarray}
The variation relative to $\zeta$ for the second form of the action
(\ref{action-2}) gives the Hamiltonian constraint
\begin{eqnarray}
 {\cal H} \equiv
 (5A-B')+\frac{1}{\zeta^2}(3E-F')+\frac{1}{\zeta^4}(H-J')-\frac{2}{\ell^2}\frac{1}{\zeta^6}
 =0 \,,
\end{eqnarray}
where
\begin{eqnarray}
 B' &=& \frac{\xi}{12\mu^4} \left\{ -\frac{3}{2}(\vec{X}'^2)^3
         - \frac{3}{2}(\vec{X}'^2)^2(\vec{X}\cdot\vec{X}'')
         - 6(\vec{X}'^2)(\vec{X}\cdot\vec{X}')(\vec{X}'\cdot\vec{X}'')  \right.   \nonumber\\
    &&  \left. +36(\vec{X}'^2)(\vec{X}\wedge\vec{X}'')^2
        + 72(\vec{X}\cdot\vec{X}'')(\vec{X}\wedge\vec{X}'')^2
        +48(\vec{X}\cdot\vec{X}')
        (\vec{X}\wedge\vec{X}'')\cdot(\vec{X}\wedge\vec{X}'')'
        \right.     \nonumber\\
    &&  \left. + 72(\vec{X}'\cdot\vec{X}'')(\vec{X}\wedge\vec{X}')\cdot(\vec{X}\wedge\vec{X}'')
        + 48(\vec{X}\cdot\vec{X}''')(\vec{X}\wedge\vec{X}')\cdot(\vec{X}\wedge\vec{X}'')
        \right.     \nonumber\\
    &&  \left. + 12(\vec{X}'^2)(\vec{X}\wedge\vec{X}')\cdot(\vec{X}\wedge\vec{X}'')'
        + 48(\vec{X}\cdot\vec{X}'')(\vec{X}\wedge\vec{X}')\cdot(\vec{X}\wedge\vec{X}'')'
        \right\}   \,,  \\
 F' &=& \frac{1}{m^2} \left\{ (\vec{X}\wedge\vec{X}'')^2
        + (\vec{X}\wedge\vec{X}')\cdot(\vec{X}\wedge\vec{X}''')
        + \frac{3}{2}(\vec{X}\wedge\vec{X}')\cdot(\vec{X}'\wedge\vec{X}'')
        \right.     \nonumber\\
    &&  \left.  - \frac{1}{4}(\vec{X}'^2)^2
        - \frac{3}{4}(\vec{X}'^2)(\vec{X}\wedge\vec{X}'') \right\}
        \,,   \\
 J' &=& -2\sigma\left\{(\vec{X}'^2)
        + (\vec{X}\cdot\vec{X}'')\right\}    \,.
\end{eqnarray}
Then the Hamiltonian constraint can be represented as follows
\begin{eqnarray}\label{HC-d}
 {\cal H} &=& \frac{\xi}{12 \mu^4} \left\{ -\frac{3}{8}(\vec{X}'^2)^3
              - 6(\vec{X}'^2)(\vec{X}\wedge\vec{X}'')^2
              - 18(\vec{X}'^2)(\vec{X}\wedge\vec{X}')\cdot(\vec{X}'\wedge\vec{X}'')
              \right.  \nonumber\\
          &&  \left.
              - 12(\vec{X}'^2)(\vec{X}\wedge\vec{X}')\cdot(\vec{X}\wedge\vec{X}''')
              - 72(\vec{X}'\cdot\vec{X}'')(\vec{X}\wedge\vec{X}')\cdot(\vec{X}\wedge\vec{X}'')
              \right.  \nonumber\\
          &&  \left.
              + 48(\vec{X}\cdot\vec{X}'')(\vec{X}\wedge\vec{X}'')^2
              - 48(\vec{X}\cdot\vec{X}''')(\vec{X}\wedge\vec{X}')\cdot(\vec{X}\wedge\vec{X}'')
              \right.   \nonumber\\
          &&  \left. - 48(\vec{X}\cdot\vec{X}')(\vec{X}\wedge\vec{X}'')\cdot(\vec{X}\wedge\vec{X}'')'
              - 48(\vec{X}\cdot\vec{X}'')(\vec{X}\wedge\vec{X}')\cdot(\vec{X}\wedge\vec{X}'')'
              \right\}  \nonumber\\
          &&  + \frac{1}{\zeta^2 m^2} \left\{ \frac{1}{2}(\vec{X}\wedge\vec{X}'')^2
              - \frac{1}{32}(\vec{X}'^2)^2 - (\vec{X}\wedge\vec{X}')\cdot(\vec{X}\wedge\vec{X}''')
              \right.  \nonumber\\
          &&  \left.
              - \frac{3}{2}(\vec{X}\wedge\vec{X}')\cdot(\vec{X}'\wedge\vec{X}'')
              \right\} + \frac{1}{\zeta^4} \frac{1}{2}\sigma(\vec{X}'^2)
              - \frac{2}{\ell^2 \zeta^6} = 0  \,.
\end{eqnarray}
In the Born-Infeld type case, the action is given by the form
(\ref{BI-action}) and the determinant of matrix ${\cal
A}^{\mu}_{~\nu} = \delta^{\mu}_{~\nu} +
\frac{\sigma}{m^2}G^{\mu}_{~\nu}$ is represented as a form
\begin{eqnarray}\label{det1}
 \det{\cal A} = 1 + \frac{\sigma}{m^2}F + \frac{1}{m^4}G
      + \frac{\sigma}{m^6}H   \,,
\end{eqnarray}
where
\begin{eqnarray}
 F &=& \zeta^2(\vec{X}\cdot\vec{X}')' - \frac{1}{4}\zeta^2(\vec{X}'^2)
       + \zeta\zeta'(\vec{X}\cdot\vec{X}')   \,, \\
 G &=& \frac{1}{4}\zeta^4 [(\vec{X}\cdot\vec{X}')'^2 - (\vec{L}'^2)]
       - \frac{1}{16}\zeta^4(\vec{X}'^2)^2
       + \frac{1}{2}\zeta^3 \zeta'
       [(\vec{X}\cdot\vec{X}')(\vec{X}\cdot\vec{X}')' - (\vec{L}\cdot\vec{L}')]  \nonumber\\
   &&  + \frac{1}{4}\zeta^2\zeta'^2[(\vec{X}\cdot\vec{X}')^2 - (\vec{L}^2)] \,, \\
 H &=& \frac{1}{64}\zeta^6(\vec{X}'^2)^3
       + \frac{1}{16}\zeta^6(\vec{X}'^2) [(\vec{X}\cdot\vec{X}')'^2
       - (\vec{X}'^2)(\vec{X}\cdot\vec{X}')' -(\vec{L}'^2)] \nonumber\\
   &&  - \frac{1}{16}\zeta^5\zeta'(\vec{X}'^2)^2(\vec{X}\cdot\vec{X}')
       + \frac{1}{8}\zeta^5\zeta'(\vec{X}'^2)
       [(\vec{X}\cdot\vec{X}')(\vec{X}\cdot\vec{X}')' - (\vec{L}\cdot\vec{L}')] \nonumber\\
   &&  + \frac{1}{16}\zeta^4\zeta'^2(\vec{X}'^2)
       [(\vec{X}\cdot\vec{X}')^2 - (\vec{L}^2)]  \,.
\end{eqnarray}
The variation of the BI type action relative to $\vec{X}$ for
equation of motion is given by
\begin{eqnarray}
 \delta{\cal S} &\simeq& \int d\rho \frac{1}{8\zeta}(\det{\cal A})^{-5/2}
      \left[ 4(\det{\cal A})^2(\vec{S}'' - \vec{Q}' + \vec{P})
      - 4(\det{\cal A})(\det{\cal A}')\vec{S}'    \right.\nonumber\\
  &&  \left. - 2(\det{\cal A})(\det{\cal A}'')\vec{S}
      + 3(\det{\cal A}')^2 \vec{S}
      + 2(\det{\cal A})(\det{\cal A}') \vec{Q} \right]
      \cdot \delta\vec{X}
\end{eqnarray}
From this variation of action, we can obtain the equation of motion
\begin{eqnarray}
 4(\det{\cal A})^2(\vec{S}'' - \vec{Q}' + \vec{P})
      - 4(\det{\cal A})(\det{\cal A}')\vec{S}'
      - 2(\det{\cal A})(\det{\cal A}'')\vec{S}  \nonumber\\
      + 3(\det{\cal A}')^2 \vec{S}
      + 2(\det{\cal A})(\det{\cal A}') \vec{Q}
      = 0   \,,
\end{eqnarray}
where
\begin{eqnarray}
 \vec{P} &=& \sigma\frac{\zeta^2}{m^2}\vec{X}''
       + \frac{\zeta^4}{m^4} \left[ \frac{1}{2}(\vec{X}'^2)\vec{X}''
       + \frac{1}{2}(\vec{X}\cdot\vec{X}')\vec{X}''
       + \frac{1}{2}(\vec{X}''^2)\vec{X}   \right.  \nonumber\\
    && \left. - \frac{1}{2}(\vec{X}\cdot\vec{X}'')\vec{X}'' \right]
       + \sigma \frac{\zeta^6}{m^6}
       \left[ \frac{1}{16}(\vec{X}'^2)^2 \vec{X}''
       + \frac{1}{8}(\vec{X}'^2)(\vec{X}\cdot\vec{X}'')\vec{X}''
       \right.  \nonumber\\
    && \left. + \frac{1}{8}(\vec{X}'^2)(\vec{X}''^2)\vec{X}
       - \frac{1}{8}(\vec{X}'^2)(\vec{X}\cdot\vec{X}'')\vec{X}''
       \right]  \,, \\
 \vec{Q} &=& \sigma \frac{\zeta^2}{m^2} \frac{3}{2}\vec{X}'
       + \frac{\zeta^4}{m^4} \left[ \frac{3}{4}(\vec{X}'^2)\vec{X}'
       + (\vec{X}\cdot\vec{X}'')\vec{X}' \right]
       + \sigma \frac{\zeta^6}{m^6}
       \left[ \frac{3}{32}(\vec{X}'^2)^2\vec{X}'  \right.  \nonumber\\
    && \left. + \frac{1}{4}(\vec{X}'^2)(\vec{X}\cdot\vec{X}'')\vec{X}'
       + \frac{1}{8}(\vec{X}\cdot\vec{X}'')^2\vec{X}'
       - \frac{1}{8}(\vec{X}\wedge\vec{X}'')^2\vec{X}'\right]  \,, \\
 \vec{S} &=& \sigma \frac{\zeta^2}{m^2} \vec{X}
       + \frac{\zeta^4}{m^4} \left[ \frac{1}{2}(\vec{X}'^2)\vec{X}
       + \frac{1}{2}(\vec{X}^2)\vec{X}'' \right]
       + \sigma \frac{\zeta^6}{m^6} \left[ \frac{1}{16}(\vec{X}'^2)^2\vec{X}
       \right. \nonumber\\
    && \left. + \frac{1}{8}(\vec{X}'^2)(\vec{X}\cdot\vec{X}'')\vec{X}
       + \frac{1}{8}(\vec{X}^2)(\vec{X}'^2)\vec{X}''
       - \frac{1}{8}(\vec{X}'^2)(\vec{X}\cdot\vec{X}'')\vec{X}
       \right]
\end{eqnarray}
with $\zeta = const$. \\
The variation of the action with $\zeta$ is given by
\begin{eqnarray}
 \delta_{\zeta}{\cal S}
  &\sim& \int d\rho \frac{1}{2\zeta^2}(\det{\cal A})^{-1/2}
        \left\{  {\cal M}
            + \frac{1}{2}\frac{(\det{\cal A}')}{\det{\cal A}}
            {\cal N} - {\cal N}' \right.  \nonumber \\
  &&        \left. - 2 \left[ (\det{\cal A})
            - (\det{\cal A})^{1/2} \left( 1+\frac{1}{2m^2\ell^2} \right) \right]
        \right\}  \delta\zeta   \,.
\end{eqnarray}
From the above variation of the action, we can get the Hamiltonian
constraint
\begin{eqnarray}
 \frac{1}{2}\frac{(\det{\cal A}')}{\det{\cal A}}{\cal N}
     + {\cal M} - {\cal N}' - 2(\det{\cal A})
     + 2(\det{\cal A})^{1/2} \left( 1+\frac{1}{2m^2\ell^2} \right)
     = 0  \,,
\end{eqnarray}
where
\begin{eqnarray}
 {\cal M} &=& \sigma\frac{\zeta^2}{m^2} \left[ 2(\vec{X}\cdot\vec{X}')'
          - \frac{1}{2}(\vec{X}'^2) \right]
          + \frac{\zeta^4}{m^4} \left[ (\vec{X}\cdot\vec{X}')'^2
          - \frac{1}{4}(\vec{X}'^2)^2
          -(\vec{L}'^2)  \right]    \nonumber\\
      &&  + \sigma\frac{\zeta^6}{m^6} \left[ \frac{3}{32}(\vec{X}'^2)^3
          + \frac{3}{8}(\vec{X}'^2)(\vec{X}\cdot\vec{X}')'^2
          - \frac{3}{8}(\vec{X}'^2)^2(\vec{X}\cdot\vec{X}')'
          - \frac{3}{8}(\vec{X}'^2)(\vec{L}'^2) \right] \,, \label{Meq}\\
 {\cal N} &=& \sigma\frac{\zeta^2}{m^2}(\vec{X}\cdot\vec{X}')
          + \frac{\zeta^4}{m^4}
          \left[ \frac{1}{2}(\vec{X}\cdot\vec{X}')
          (\vec{X}\cdot\vec{X}')'
          - \frac{1}{2}(\vec{L}\cdot\vec{L}')
          \right]  \nonumber\\
     &&   + \sigma\frac{\zeta^6}{m^6} \left[
          - \frac{1}{16}(\vec{X}'^2)^2(\vec{X}\cdot\vec{X}')
          + \frac{1}{8}(\vec{X}'^2)(\vec{X}\cdot\vec{X}')
          (\vec{X}\cdot\vec{X}')'
          - \frac{1}{8}(\vec{X}'^2)(\vec{L}\cdot\vec{L}') \right]
          \label{Neq}
\end{eqnarray}
with $\zeta = const$.

\newpage
%%%%%%%%%%%%%%%%%%%%%%%%%%%%%%%%%%%%%%%%%%%%%%%%%%%%%%%%%%%%%%%%%%%%%

\end{document}